# Eco-Innovation and Earnings Management: Unveiling the Moderating Effects of Financial Constraints and Opacity in FTSE All-Share Firms

Probowo Erawan Sastroredjo[1,2]*, Marcel Ausloos [3,4,5,6] and Polina Khrennikova [7,8]

[1] University of Leicester, Leicester, United Kingdom
[2] Parahyangan Catholic University, Bandung, Indonesia
[3] University of Leicester, Leicester, United Kingdom,
[4] Group of Researchers Applying Physics in Economy and Sociology (GRAPES), Liège, Belgium
[5] Babeș-Bolyai *University*, Cluj-Napoca, Romania
[6] Bucharest University of Economic Studies, Bucharest, Romania
[7] University of Leicester, Leicester, United Kingdom
[8] University of Twente, Enschede, The Netherlands
*Corresponding author: pes13@leicester.ac.uk

**Abstract:** Our research investigates the relationship between eco-innovation and earnings management among 567 firms listed on the FTSE All-Share Index from 2014 to 2022. By examining how sustainability-driven innovation influences financial reporting practices, we explore the strategic motivations behind income smoothing in firms engaged in environmental initiatives. The findings reveal a positive association between eco-innovation and earnings management, suggesting that firms may leverage eco-innovation not only for environmental signalling but also to project financial stability and meet stakeholder expectations. The analysis further uncovers that the propensity for earnings management is amplified in firms facing financial constraints, proxied by low Whited-Wu (WW) scores and weak sales performance, and in those characterised by high financial opacity. We employ a robust multi-method approach to address potential endogeneity and selection bias, including entropy balancing, propensity score matching (PSM), and the Heckman Test correction. Our research contributes to the literature by providing empirical evidence on the dual strategic role of eco-innovation—balancing sustainability signalling with earnings management—under varying financial conditions. The findings offer actionable insights for regulators, investors, and policymakers navigating the intersection of corporate transparency, financial health, and environmental responsibility.

## 1. Introduction

Earnings management is an important area of finance and corporate governance research that shows how managers influence reported financial outcomes to attain goals such as benchmark beating or masking poor performance. Several earlier studies, such as Healy and Wahlen [1] and Dechow et al. [2], discussed the motivations and mechanisms behind earnings management. Various identified factors include financial constraints, reputational concerns, and regulatory compliance. However, despite the vast literature on earnings management, a critical gap remains in understanding how it intersects with emerging corporate sustainability practices, especially those involving eco-innovation. This study, therefore, seeks to fill this gap by examining whether engagement in eco-innovation activities affects a firm's propensity for earnings management. More precisely, the research attempts to answer the question: Does engagement in eco-innovation encourage a firm to engage in earnings management practices?

Eco-innovation has earned significant interest among academics and policymakers due to the increasing global focus on ESG. The UK is one of the forerunners in this regard, with policies in tune with the United Nations sustainability goals to ensure that corporate eco-innovation becomes a source of competitive advantage. While previous literature has addressed different dimensions of eco-innovation, such as its technological drivers [3], market influences [4], and regulatory impacts [5], not many have explored its financial repercussions, especially concerning earnings management. Based on these factors, our research centres on UK firms, as they offer a rich context for examining the relationship between eco-innovation and financial reporting, given their advanced regulatory framework, strong ESG ecosystem, and diverse industrial landscape. Several theoretical perspectives help explain how earnings management and eco-innovation are related. The technology push theory suggests that firms engaging in eco-innovation invest heavily in research and development [6], incurring significant financial pressures. Such economic strain might prompt managers to manipulate earnings to maintain financial stability or meet market expectations [2, 7]. Based on the discussion, we propose a positive relationship between a firm's eco-innovation level and its tendency toward earnings management practices.

Our research employs 567 UK firms listed in the FTSE All-Share index between 2014 and 2022, using data obtained from Refinitiv Eikon. A robust OLS regression model is employed to investigate the effect of eco-innovation on earnings management. The main variables are eco-innovation and earnings management, which were estimated using prior literature. The findings from our research show a significantly positive relationship between eco-innovation and earnings management, hence supporting the hypothesis.

Several robustness tests and checks for endogeneity are conducted to ensure that potential endogeneity issues do not drive the results. To overcome some of the biases in the analysis, this study employs entropy balancing, PSM, and the Heckman Test model. Alternative measures of earnings management, such as those developed by Jones [8], Dechow [9], Kasznik [10], and Kothari [11], are employed in the study to ensure the robustness of the findings. These tests confirm the consistency of the observed relationship between eco-innovation and earnings management, strengthening the reliability of the study's conclusions. We also examine the moderating effects of financial constraints and opacity on the relationship between eco-innovation and earnings management. These analyses show that the observed relationship between eco-innovation and earnings management is more pronounced among financially constrained firms and less pronounced for firms with low opacity.

Our research makes several significant contributions to the literature. First, it provides novel insights into the intersection between eco-innovation and earnings management, a relatively unexplored research area. Second, it deepens the understanding of how eco-innovation, as an integral part of ESG practices, affects financial reporting behaviours. Our study develops a fine-grained view of how eco-innovation shapes earnings management by including financial constraints and opacity as moderating variables. This will be realised through a strong methodological framework that provides credible and actionable insights into these issues by academics, policy analysts, and sustainability and corporate governance practitioners.

This paper is subdivided into several sections. Section 1 introduces the topic. Section 2 presents a review of the literature and the development of the hypothesis. Section 3 discusses the research design and the data utilised, while Section 4 analyses the regression results and the tests for endogeneity. Finally, Section 5 discusses the conclusions derived from the findings, including the limitations of our research and recommendations for future research.

## 2. Literature Review and Hypothesis Development

*2.1. Understanding Earnings Management*

Scholars have defined and interpreted earnings management differently to suit various research projects. According to Stickney et al. [12], earnings management involves manipulating financial reports to gain personal benefits. This is consistent with Schipper's views [13]. The most common aim of this practice is to provide stable and reliable income reports that indicate the company's consistent performance [14].

Interestingly, Graham et al. [15] reveal that many corporate executives prefer to manage earnings rather than report volatile income streams, as investors and analysts often perceive income instability as a sign of financial weakness and managerial inefficiency. This preference underscores the managerial incentives tied to market expectations and the short-termism prevalent in corporate reporting culture. Supporting this view, the agency cost theory, advanced by Jensen and Meckling [16], posits that managers, acting in their own interests rather than shareholders', may manipulate earnings by deflating previously inflated profits to mitigate future scrutiny and risks. Boachie and Mensah [14] further argue that such behaviour is often rationalised as a form of risk management under information asymmetry.

Scholars have increasingly examined the intersection between earnings management and corporate innovation in recent years. However, the empirical evidence remains mixed. While some studies, such as Jia and Lin [17], argue that earnings management inhibits innovation by diverting resources away from long-term investment in R&D, others have found that it may, paradoxically, stimulate innovation—albeit selectively. For instance, Chouaibi et al. [18] document that innovative firms with higher R&D intensity are more likely to engage in upward real earnings management, suggesting a positive association in general industrial contexts. Conversely, Zhang [19] finds that managers in firms with equity incentive plans may manipulate R&D investment intensity as a form of real earnings management, which can have varying impacts depending on the industry and firm characteristics.

Despite this growing body of literature, a significant gap exists. Yang et al. [20] highlight the lack of empirical research specifically addressing the impact of earnings management on environmentally friendly innovation, commonly referred to as eco-innovation. Given the rising importance of sustainability and environmental accountability, it is crucial to understand how managerial financial behaviour influences firms' commitment to green innovation. Therefore, our research aims to fill this gap by analysing the relationship between earnings management and eco-innovation among firms listed in the UK, focusing on the FTSE All-Share Index.

*2.2. Previous studies on Eco-Innovation*

In recent years, businesses have increasingly focused on environmental innovation and embraced socially responsible practices [21]. This positive trend aligns with the growing importance of ESG initiatives, which gained momentum after the introduction of sustainable development at the UN's 1992 Rio Summit [22].

Eco-innovation, also known as environmental, green, or sustainability innovation, refers to advancements that foster environmental sustainability through ecological improvements [23, 24, 25, 26]. According to Demirel and Kesidou [27], eco-innovation can be categorised by its impact levels: End-of-Pipeline Pollution Control Technologies, implemented at the final production stage, manage emissions and waste without modifying the core production process. Despite their effectiveness, these solutions are frequently considered costly and classified as substantial investments [7, 27, 28]. Integrated Cleaner Production Technologies aim to minimise pollution by modifying production processes or using eco-friendly materials. Unlike end-of-pipe solutions, these technologies embed environmental protection into manufacturing, reducing raw material usage, energy consumption, and regulatory compliance costs [27, 29].

Environmental R&D, on the other hand, focuses on enhancing processes and products to achieve cleaner production and consumption. While these innovations often yield significant environmental benefits, they involve higher risks and costs [27]. These categories illustrate eco-innovation's diverse purposes and implementation strategies, highlighting its pivotal role in advancing sustainability goals.

Recent literature has begun exploring the dual-use nature of eco-innovation, where sustainability commitments intersect with financial reporting concerns. Almubarak et al. [30] underscore that firms possessing active ESG programs, particularly those experiencing financial distress, may resort to earnings management to sustain perceptions of stability. Likewise, Albitar et al. [31] emphasise that eco-innovation initiatives are frequently correlated with robust climate governance; however, the financial pressures associated with such commitments may inadvertently generate incentives for earnings management.

Furthermore, Amin et al. [32] and Jiang et al. [33] indicate that companies may use ESG practices to disguise aggressive tax avoidance or unclear financial operations when uncertain climate policy. This is consistent with the findings of Souguir et al. [34], who express scepticism regarding whether green disclosures genuinely reflect legitimate efforts or are sometimes employed as greenwashing tactics, thereby emphasising concerns related to financial transparency and stakeholder trust misperceptions. These recent contributions reinforce the importance of examining eco-innovation not in isolation, but as part of a broader strategic narrative that includes financial motivations and constraints. As our study explores, understanding this interplay is essential to assess how firms manage earnings, navigate transparency obligations, and maintain legitimacy under increasing ESG scrutiny.

*2.3. Theoretical Underpinning*

Several theories explain why companies adopt eco-innovation. First is the "technology push effect," hypothesising that through R&D activities, companies innovate due to technological advances and availability [6]. Consequently, there are increased capabilities toward using technologies with a way forward to produce innovative products or processes [35]. Another perspective is the "market pull" theory, which states that eco-innovation results from increased awareness of environmental sustainability among the public. As such, companies try to meet this growing demand by developing eco-friendly products to satisfy market expectations [3]. The third theory, the "firm-specific factor" theory, suggests that even though research and development may increase a company's capacity, its innovation is restricted by its core business operations and priorities [36]. The "regulation push/pull" theory explains that stringent environmental regulations are essential in fostering innovation. The more environmentally binding policies are, the higher the motivation for businesses to indulge in environmental research and development to drive eco-innovation [3, 37, 38, 39, 40].

ESG practices and eco-innovation serve as fundamental instruments for acquiring and maintaining legitimacy [41, 42, 43]. Eco-innovation involves the development of new or enhanced products, processes, or practices aimed at reducing environmental harm while underscoring a company's proactive commitment to sustainability [44]. By actively participating in ESG initiatives or pursuing eco-innovation, companies comply with regulatory requirements and align with societal values. This alignment mitigates reputational risks and bolsters their credibility, financial stability, and legitimacy [43, 45, 46, 47].

Despite the potential benefits, implementing eco-innovation is not without significant challenges. From the agency theory perspective, managers may pursue eco-innovation initiatives to project a positive public image or satisfy external stakeholders, even when such projects are not aligned with shareholder interests or do not contribute to firm value. This divergence can lead to inefficiencies and opportunistic behaviour, especially when monitoring mechanisms are weak. Additionally, cash flow theory suggests that firms facing internal financing constraints may struggle to support the often-high upfront costs associated with eco-innovation [48]. These financial limitations may force firms to prioritise short-term liquidity over long-term sustainability investments [49], potentially create a deficiency of transparency in their information disclosures [50, 51] and increase the risk of earnings management to mask performance fluctuations [52].

While the technology-push, market pull, firm-specific factor, and regulation-pull theories explain eco-innovation motivations, Alternative frameworks such as legitimacy theory provide insights into the significance of implementing eco-innovation for companies' sustainability. Conversely, the application of eco-innovation, as emphasised by agency theory and cash flow theory, may pose challenges for companies due to potential costs incurred. This dual perspective facilitates contextualisation of our findings within ethical and strategic frameworks for implementing eco-innovation.

*2.4. Hypothesis Development*

In this section, we build upon the theoretical framework introduced by investigating three key theories that connect earnings management with eco-innovation as we formulate our research hypotheses. We first emphasise the significance of eco-innovation in fostering community relations through the legitimacy theory perspective. Furthermore, we analyse possible financial obstacles associated with implementing eco-innovation, particularly the financial limitations highlighted in cash flow theory, which could lead to reduced financial transparency, as noted in agency theory.

*2.4.1. Eco-Innovation and Earnings Management*

Firms must conduct extensive research and strategic planning to pinpoint the most promising innovation opportunities, which is crucial due to the significant investment costs [53]. However, the funding required may not always align with the company's financial capabilities. Firms that pursue eco-innovation often face significant financial challenges due to the substantial investments required for research and development [6]. The financial strain from these activities can pressure managers to ensure the firm's financial stability, especially when external stakeholders and investors expect consistent profitability [2, 7]. Firms, therefore, respond by engaging in earnings management to portray the appearance of good and stable financial performance even in the most adverse economic times. In this regard, smoothing earnings may help firms dampen the negative implications of financial turbulence on stakeholder confidence.

Besides, eco-innovation is not only a financial but also a reputational commitment. Usually, firms that adopt environmentally innovative practices are pressured to match sustainability objectives with good financial performance [54]. Not being able to project continuous financial success hand in hand with eco-innovation efforts may blemish their reputation and undermine stakeholder trust. To meet these reputational demands, managers may use earnings management as a strategic approach to ensure that financial outcomes appear favourable. By doing so, firms balance their environmental and financial goals, presenting a cohesive narrative to stakeholders.

In addition, eco-innovation stands out as an especially delicate balance for firms aiming to pursue environmental ends while maintaining wholesome financial performance; such firms, therefore, can be said to face double pressures-economic burdens ensuing from high R&D expenditure [6], and finding it necessary to maintain their reputation for being innovative and lucrative [54]. Earnings management, in this stern environment, helps balance these contradicting needs. As a strategic move, management may seek financial stability, stakeholders' expectations, and confirmation of the firm's commitment to sustainability without revealing its financial vulnerability through income smoothing. Based on the discussion above, this paper hypothesises that:

**H1:** Firms with higher levels of eco-innovation engage in more earnings management practices.

*2.4.2. The Moderating Effect of Financial Constraint*

According to cash flow theory, a stable cash flow is crucial for a company's growth and sustainability [52]. Yet, forecasting cash inflows is inherently uncertain. Firms need operational funds, particularly for eco-innovation, which can be expensive [27]. Therefore, additional funding sources are essential to meet the expectations and demands of eco-innovation implementation. Financial limitations hinder access to these funds, increasing the risk of manipulation. As a result, this study hypothesises that financial constraints heighten the likelihood of a firm engaging in earnings management while pursuing eco-innovation.

**H2:** The positive effect of eco-innovation on earnings management is stronger for firms with high financial constraints.

*2.4.3. The Moderating Effect of Financial Opacity*

Financial opacity has been considered one of the major determinants of financial manipulation [55, 56]. On the other hand, higher transparency in financial information has been shown to enhance management accountability in financial reporting, thus reducing earnings management practices [57]. Based on such insights, this paper argues that low levels of transparency create an environment where earnings management practices can thrive. This relationship becomes pronounced as companies pursue innovation in environments where pressures to meet performance expectations could intensify managerial incentives to manage earnings.

**H3:** The positive effect of eco-innovation on earnings management is stronger for firms with high financial opacity (lack of financial transparency).

## 3. Research Design and Data

*3.1. Sample Selection*

The sample of firms is selected from those listed in the FTSE All-Share Index. Eco-innovation and financial data for these firms were sourced from the Refinitiv Eikon Database for nine years, from 2014 to 2022, with a total of 1,878 firm-year observations. We relied on Refinitiv Eikon for its comprehensive ESG and financial reporting data, which are widely used in academic research. The eco-innovation score is derived from ESG subcomponents, ensuring standardised and comparable measurement across firms.

The period selected is consistent with that used by Albitar et al. [31]. Our research excludes firms belonging to the finance and utility industries, as these have peculiarities in financial management and reporting [51, 58]. After the exclusion criteria, the final sample consists of 1,609 firm-year observations across 567 firms. This study also relies on the Fama-French 12 Industry Classification in conducting its analysis. The sample also reflects the diversity of sectors within the FTSE All-Share Index, enhancing the generalizability of our findings to the market contexts. All continuous variables are winsorised at 1% and 99% to lessen the effect of outliers.

**Table 1.** Sample selection.

| Criteria | Number of Firm-Years |
|---|---|
| Refinitiv Eikon | 1,878 |
| Less: | |
| Utility Firms (SIC codes 6000-6999) | 60 |
| Financial Firms (SIC codes 4000-4999) | 200 |
| Others (SIC codes between 4000-4999 and 6000-6999) | 9 |
| Final sample | 1,609 |

*3.2. Variable Measurement*

*3.2.1. Dependent variable*

Following prior literature, we use discretionary accruals (*DACC*) as a proxy for earnings management [11, 30]. *DACC* captures the discretionary component of total accruals, encompassing both opportunistic and potentially informative earnings adjustments. However, consistent with the dominant view in sustainability and ESG research [31, 34], we interpret higher discretionary accruals as indicative of opportunistic earnings management, particularly in contexts characterised by financial opacity or reputational pressures. While this approach does not allow us to isolate opportunism per se, it is widely adopted in the earnings management literature as a reasonable proxy for such behaviour. The calculation method for *DACC* is as follows [11]:

$$\frac{TA_{it}}{A_{it-1}} = \alpha_0 + \alpha_1 \left[\frac{1}{A_{it-1}}\right] + \alpha_2 \left[\frac{\Delta REV_{it} - \Delta REC_{it}}{A_{it-1}}\right] + \alpha_3 \left[\frac{PPE_{it}}{A_{it-1}}\right] + \alpha_4 \left[\frac{ROA_{it}}{A_{it-1}}\right] + \varepsilon_{it} \quad (1)$$

Where $TA_{it}$ is total accrual in year *t* for firm *i*, $A_{it-1}$ is total assets in year *t* – 1 for firm *i*, $\Delta REV_{it}$ is revenues in year *t* less revenue in year *t* - 1 for firm *i*, $\Delta REC_{it}$ is receivable in year *t* less receivable in year *t* - 1 for firm *i*, $PPE_{it}$ is gross property, plant, and equipment in year *t* for firm *i*, $ROA_{it}$ is a return on assets in year *t* less receivable in year *t* - 1 for firm *i*.

*3.2.2. Independent variable*

The key variable of interest in our research is the eco-innovation score (*EIS*). This variable assesses the degree to which the company prioritises eco-innovation elements within the ESG framework and its potential impact on earnings management strategies. *EIS* is estimated by dividing the Eco-innovation Score by 100.

### 3.2.3. Control variables

A set of control variables has been included to ensure the robustness and validity of the results obtained in this research. The control variables utilised in this study include the following: Return on assets (*ROA*) is calculated as the income after taxes for the fiscal period divided by the average total assets [31, 33, 59]. The natural logarithm of total assets [60] determines the company's size (*SIZE*). Leverage (*LEV*) is calculated as the total debt divided by total assets [33, 34, 59, 61]. The market-to-book ratio (*MTB*) is measured by dividing the company's market capitalisation by the book value [32, 59]. Asset-income ratio (*AIN*) is determined by dividing total assets by net income before taxes [33]. The liquidity ratio (*LIQ*) is calculated by dividing current assets by current liabilities [31]. Firm age (*AGE*) is measured by the natural logarithm of year t minus the date of incorporation plus 1 [33]. Big 4 (*BIG4*) is calculated as a dummy variable to denote whether the auditor is affiliated with one of the Big 4 auditor firms (1) or not (0) [59, 62, 63]. The natural logarithm of the number of board members measures board size (*BODSIZE*) [31, 64]. Board independence (*BODIND*) is measured by the proportion of independent directors on the board [31, 64]. Capital intensity (*CAP_INTENS*) is measured by firm capital divided by total sales [31].

**Table 2.** Variables Definition.

| **Dependent Variables** | Description | Source |
|---|---|---|
| Discretionary accruals *(DACC)* | a proxy to represent earnings management, using the Kothari model | Refinitiv Eikon |
| **Independent Variable** | | |
| Eco-innovation Score *(EIS)* | reflects a company's capacity to reduce the environmental costs and burdens for its customers and thereby creating new market opportunities through new environmental technologies and processes or eco-designed products. The range score is between 0 – 100 | Refinitiv Eikon |
| **Control Variables** | | |
| Return on assets *(ROA)* | measured as income after taxes for the fiscal period divided by the average total assets | Refinitiv Eikon |
| Size *(SIZE)* | measured by the natural logarithm of total assets | Refinitiv Eikon |
| Leverage *(LEV)* | measured by total debt divided by total assets | Refinitiv Eikon |
| Market-to-book ratio *(MTB)* | measured by company market capitalisation divided by book value capitalisation | Refinitiv Eikon |
| Asset-income ratio *(AIN)* | measured by total assets divided by net income before taxes | Refinitiv Eikon |
| Liquidity *(LIQ)* | measured as current assets divided by current liabilities | Refinitiv Eikon |
| Firm age *(AGE)* | measured by the natural logarithm of year t minus the date of incorporation plus 1 | Refinitiv Eikon |
| Big 4 *(BIG4)* | a dummy variable to denote whether the auditor is affiliated with one of the BIG 4 auditor firms (1) or not (0) | Refinitiv Eikon |
| Board size *(BODSIZE)* | measured by the natural logarithm of the number of board members | Refinitiv Eikon |
| Board independence *(BODIND)* | measured by the proportion of independent directors on the board | Refinitiv Eikon |
| Capital Intensity *(CAP_INTENS)* | Measured by firm capital divided by total sales | Refinitiv Eikon |

*3.3. Regression Model Estimation*

We build the model upon Chouaibi et al. [18] research model by refining earnings management calculations, incorporating the eco-innovation score as an independent variable, and integrating several control variables linked to earnings management. We conduct a sequence of tests on the model to explore how the independent variable affects the dependent variable. Initially, we analyse this relationship without control variables or fixed effects (refer to Eq. 2). Then, we assess the impact of the independent variable on the dependent variable with the control variable included. Still, without the fixed effect variable (see Eq. 3). Lastly, we perform an analysis that incorporates all variables (see Eq. 4). The model used to investigate this relationship is detailed below:

$$DACC_{it}^{(1)} = \beta_0^{(1)} + \beta_1^{(1)} EIS_{it} + \varepsilon_{it}^{(1)} \tag{2}$$

$$DACC_{it}^{(2)} = \beta_0^{(2)} + \beta_1^{(2)} EIS_{it} + \beta_2^{(2)} ROA_{it} + \beta_3^{(2)} SIZE_{it} + \beta_4^{(2)} LEV_{it} + \beta_5^{(2)} MTB_{it} + \beta_6^{(2)} AIN_{it} + \beta_7^{(2)} LIQ_{it} + \beta_8^{(2)} AGE_{it} + \beta_9^{(2)} BIG4_{it} + \beta_{10}^{(2)} BODSIZE_{it} + \beta_{11}^{(2)} BODIND_{it} + \beta_{12}^{(2)} CAP\_INTENS_{it} + \varepsilon_{it}^{(2)} \tag{3}$$

$$DACC_{it}^{(3)} = \beta_0^{(3)} + \beta_1^{(3)} EIS_{it} + \beta_2^{(3)} ROA_{it} + \beta_3^{(3)} SIZE_{it} + \beta_4^{(3)} LEV_{it} + \beta_5^{(3)} MTB_{it} + \beta_6^{(3)} AIN_{it} + \beta_7^{(3)} LIQ_{it} + \beta_8^{(3)} AGE_{it} + \beta_9^{(3)} BIG4_{it} + \beta_{10}^{(3)} BODSIZE_{it} + \beta_{11}^{(3)} BODIND_{it} + \beta_{12}^{(3)} CAP\_INTENS_{it} + Industry\ Fixed\ Effect + Year\ Fixed\ Effect + \varepsilon_{it}^{(3)} \tag{4}$$

*DACC* refers to the primary earnings management variable, as explained in section 3.2.1. *EIS* refers to the eco-innovation score, as described in section 3.2.2. *ROA refers to the* return on assets, *SIZE* refers to the size of the company, *LEV* refers to leverage, *MTB* refers to the market-to-book ratio, *AIN* refers to the asset-income ratio, *LIQ* refers to the liquidity ratio, *AGE* refers to the firm age (date of incorporation), *BIG4* refers to a dummy variable of the BIG 4 auditor firms, *BODSIZE* refers to the board size of FTSE All-Share companies, and *BODIND* refers to the independent director on the board at FTSE All-Share companies. All control variables are explained in section 3.2.3. Industry (based on Fama-French 12 industry classification) and Year Fixed Effects are also controlled.

*3.4. Data Analysis Strategy*

To test the proposed hypotheses, we conduct a structured, multi-stage data analysis using panel regression techniques and a suite of robustness checks. Our analysis proceeded as follows:

*3.4.1. Baseline Regression*

We begin by estimating Ordinary Least Squares (OLS) regression models to examine the effect of the eco-innovation score (*EIS*) on earnings management using *DACC*. Initially, we include only the primary variables, and subsequently, we add control variables. We also control for time and industry effects using year and industry fixed effects based on the Fama-French 12-industry classification.

*3.4.2. Moderation Analysis*

To investigate Hypotheses 2 and 3, we analyse how financial constraints and opacity moderate the results. This involves using interaction terms and conducting subsample analyses, employing proxies like the *Whited-Wu (WW) score*, firm *sales* to indicate financial constraints, and the *FFIN* transparency for financial opacity.

*3.4.3. Robustness Checks*

To validate the reliability of our findings, we employ four alternative models of earnings management: the Jones [8], Dechow et al. [9], Kasznik [10], and Kothari et al. [11] models. The consistent results across all models confirm the robustness of the eco-innovation effect on earnings management.

*3.4.4. Endogeneity and Selection Bias Corrections*

We further addressed potential endogeneity concerns using three techniques: *Entropy Balancing*, to reweight the sample and achieve covariate balance between treatment and control groups, *Propensity Score Matching (PSM)*, using one-to-one matching with a calliper (0.001), to reduce selection bias based on observable characteristics, and *Heckman Test*, to control for selection bias by modelling eco-innovation adoption in the first stage and including the inverse Mills ratio in the second-stage outcome model. This multi-method approach allows for triangulating findings across different estimation strategies, mitigating concerns over bias due to omitted variables, sample selection, or observable heterogeneity. Each technique was selected based on theoretical appropriateness and empirical validation in similar ESG-finance contexts. All models were estimated using heteroskedasticity-robust standard errors clustered at the firm level, and significance was evaluated at the 1%, 5%, and 10% levels.

**Figure 1.** Theoretical Framework.

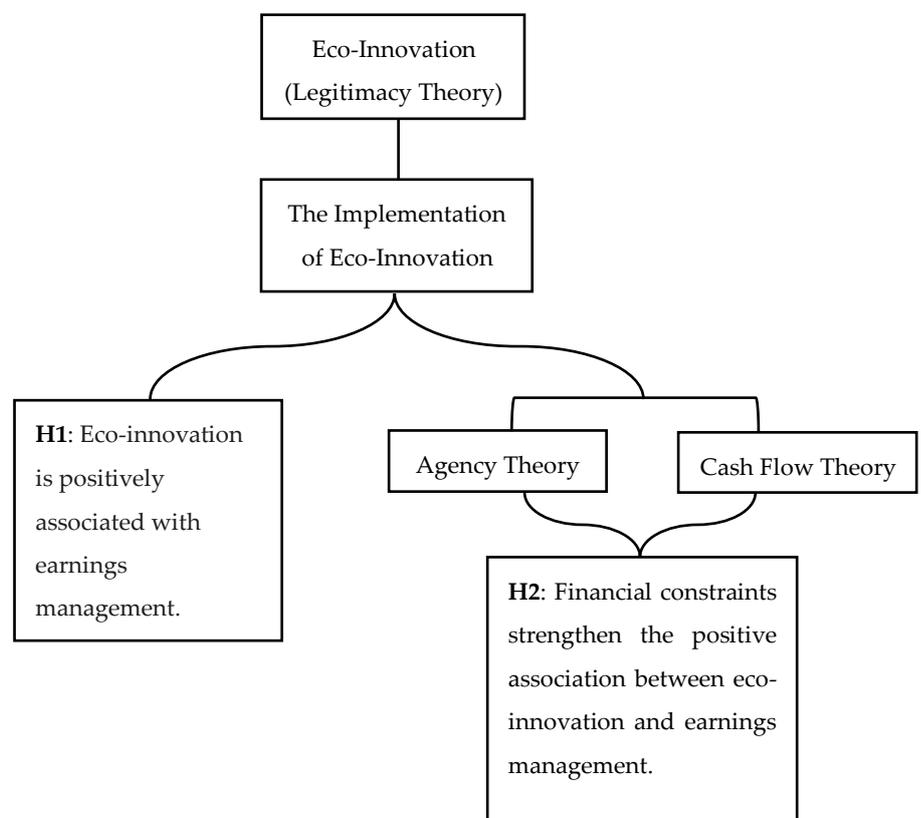

## 4. Empirical Result

*4.1. Descriptive Statistics*

*4.1.1. Distribution by industry and year*

Table 3 displays the sample distribution by year and industry. Panel A reports the distribution by year. The number of observations steadily increased, with the lowest in 2014 and the highest in 2022. Panel B also shows the distribution by industry. Notably, the consumer durables industry had the lowest observations, with only 12, while the highest observations were observed for firms in other industries, with 545.

**Table 3.** Distribution by year and Industry.

**Panel A:** Distribution by Year

| year | Freq. | Percent | Cum. |
| --- | --- | --- | --- |
| 2014 | 90 | 5.59 | 5.59 |
| 2015 | 148 | 9.20 | 14.79 |
| 2016 | 158 | 9.82 | 24.61 |
| 2017 | 170 | 10.57 | 35.18 |
| 2018 | 171 | 10.63 | 45.80 |
| 2019 | 184 | 11.44 | 57.24 |
| 2020 | 224 | 13.92 | 71.16 |
| 2021 | 227 | 14.11 | 85.27 |
| 2022 | 237 | 14.73 | 100.00 |
| Total | 1609 | 100.00 | |

**Panel B:** Distribution by Industry.

| | Freq. | Percent | Cum. |
| --- | --- | --- | --- |
| Consumer Nondurables | 143 | 8.89 | 8.89 |
| Consumer Durables | 12 | 0.75 | 9.63 |
| Manufacturing | 211 | 13.11 | 22.75 |
| Oil, Gas, and Coal Extraction and Products | 75 | 4.66 | 27.41 |
| Chemicals and Allied Products | 64 | 3.98 | 31.39 |
| Business Equipment | 156 | 9.70 | 41.08 |
| Telephone and Television Transmission | 19 | 1.18 | 42.26 |
| Whole sales, Retail, and Some Services | 340 | 21.13 | 63.39 |
| Healthcare, Medical Equipment, and Drugs | 44 | 2.73 | 66.13 |
| Other | 545 | 33.87 | 100.00 |
| Total | 1609 | 100.00 | |

*4.1.2. Summary statistics, variance inflation factors and correlation matrix*

Table 4 reports the summary distribution of all variables used in the primary analysis, variance inflation factors, and the correlation matrix. From Panel A, it is observed that the mean *DACC* is -0.006, while the mean Eco-innovation (*EIS*) is 0.265. The average profitability is around 6.2%, and the average firm size is around 21. It is observed that *BIG4* audit firms audited 64.2% of the firms in our sample.

Panel B shows the variance inflation factors for the variables used for our primary analysis. The results from the test confirm no signs of multicollinearity in the primary model, as indicated by VIF values < 10 and 1/VIF values > 0.10. As reported in Panel C, the correlation matrix reveals that the dependent variable, *DACC*, is positively correlated with several variables, particularly eco-innovation (*EIS*) and the firm's age (*AGE*).

**Table 4. Summary statistics, variance inflation factors and correlation matrix**

**Panel A:** Summary Statistics

|  | N | Mean | SD | Min | Median | Max |
|---|---|---|---|---|---|---|
| *DACC* | 1,609 | -0.006 | 0.064 | -0.314 | -0.002 | 0.321 |
| *EIS* | 1,609 | 0.265 | 0.313 | 0.000 | 0.120 | 0.986 |
| *ROA* | 1,609 | 0.062 | 0.095 | -0.248 | 0.055 | 0.448 |
| *SIZE* | 1,609 | 21.161 | 1.560 | 16.435 | 21.061 | 25.512 |
| *LEV* | 1,609 | 0.204 | 0.154 | 0.000 | 0.200 | 0.770 |
| *MTB* | 1,609 | 1.825 | 1.895 | 0.235 | 1.217 | 12.875 |
| *AIN* | 1,609 | 0.139 | 0.615 | -2.727 | 0.061 | 4.391 |
| *LIQ* | 1,609 | 1.633 | 1.187 | 0.242 | 1.333 | 8.112 |
| *AGE* | 1,609 | 3.233 | 0.977 | 0.000 | 3.180 | 4.930 |
| *BODSIZE* | 1,609 | 8.804 | 2.084 | 3.000 | 9.000 | 17.000 |
| *BODIND* | 1,609 | 0.615 | 0.130 | 0.071 | 0.625 | 0.923 |
| *BIG4* | 1,609 | 0.642 | 0.480 | 0.000 | 1.000 | 1.000 |
| *CAP_INTENS* | 1,609 | 1.366 | 2.386 | 0.057 | 0.866 | 26.120 |

**Panel B:** Variance inflation factors

| Variable | VIF | 1/VIF |
|---|---|---|
| *SIZE* | 2.20 | 0.454 |
| *MTB* | 1.70 | 0.587 |
| *ROA* | 1.68 | 0.596 |
| *BODSIZE* | 1.60 | 0.626 |
| *CAP_INTENS* | 1.32 | 0.758 |
| *LEV* | 1.26 | 0.793 |
| *LIQ* | 1.21 | 0.827 |
| *EIS* | 1.19 | 0.837 |
| *BODIND* | 1.16 | 0.865 |
| *AGE* | 1.12 | 0.889 |
| *AIN* | 1.05 | 0.949 |
| *BIG4* | 1.03 | 0.975 |
| *MEAN VIF* | 1.38 | |

**Panel C:** Correlation Matrix

| Variables | (1) | (2) | (3) | (4) | (5) | (6) | (7) | (8) | (9) | (10) | (11) | (12) | (13) |
|---|---|---|---|---|---|---|---|---|---|---|---|---|---|
| *DACC* | 1.000 | | | | | | | | | | | | |
| *EIS* | 0.096*** | 1.000 | | | | | | | | | | | |
| *ROA* | -0.035 | -0.007 | 1.000 | | | | | | | | | | |
| *SIZE* | 0.008 | 0.334*** | -0.094*** | 1.000 | | | | | | | | | |
| *LEV* | -0.080*** | -0.015 | -0.252*** | 0.145*** | 1.000 | | | | | | | | |
| *MTB* | -0.302*** | -0.085*** | 0.574*** | -0.241*** | -0.065*** | 1.000 | | | | | | | |
| *AIN* | 0.050** | -0.063** | 0.042* | -0.190*** | -0.059** | 0.030 | 1.000 | | | | | | |
| *LIQ* | 0.030 | -0.064** | 0.216*** | -0.236*** | -0.313*** | 0.105*** | 0.004 | 1.000 | | | | | |
| *AGE* | 0.165*** | 0.190*** | -0.035 | 0.143*** | -0.171*** | -0.103*** | 0.014 | 0.009 | 1.000 | | | | |
| *BODSIZE* | 0.011 | 0.218*** | -0.107*** | 0.534*** | 0.180*** | -0.092*** | -0.161*** | -0.071*** | -0.044* | 1.000 | | | |
| *BODIND* | -0.002 | 0.255*** | -0.033 | 0.298*** | 0.077*** | -0.112*** | -0.074*** | -0.061** | 0.100*** | 0.221*** | 1.000 | | |
| *BIG4* | -0.042* | -0.010 | 0.087*** | -0.011 | -0.019 | 0.120*** | 0.032 | -0.029 | 0.048* | -0.052** | 0.013 | 1.000 | |
| *CAP_INTENS* | 0.190*** | -0.081*** | -0.121*** | -0.322*** | 0.106*** | -0.109*** | -0.018 | 0.096*** | -0.143*** | 0.045* | 0.011 | -0.024 | 1.000 |

*** $p<0.01$, ** $p<0.05$, * $p<0.1$

*4.2. Eco-innovation and earnings management (Hypothesis 1)*

Table 5 reports the regression results for the primary hypothesis. In column 1, a regression is conducted for only the dependent and independent variables, with no control variables and no fixed effects. The eco-innovation score (*EIS*) coefficient is 0.195, which is significant at the 1% level and suggests a positive relationship with earnings management (*DACC*). The R-squared value is 0.009, indicating that approximately 1% of the variance in *DACC* can be explained by *EIS* alone.

Column 2 presents the regression between *DACC* and *EIS* with control variables without fixed effects. The coefficient for *EIS* is 0.152, which remains significant at the 1% level, indicating a positive relationship. *ROA*, *AGE*, and *CAP_INTENS* are also positively and significantly related at the 1% level. The R-squared value is 0.181, indicating that the model explains approximately 18.1% of the variance in *DACC*.

The regression results in column 3 include the year and Industry fixed effects. The variable *EIS* is positively associated with *DACC*, showing that as *EIS* increases, so does *DACC*. Other variables, such as *ROA*, *AGE*, *AIN*, and *CAP_INTENS*, also demonstrated positive effects, indicating that the rise of these variables influences the increase in earnings management. Overall, the results are consistent with the primary hypothesis that firms with a higher eco-innovation level engage in more earnings management practices.

**Table 5. Baseline Regression result**

| Variables | DACC | | |
|---|---|---|---|
| | (1) | (2) | (3) |
| EIS | 0.195*** (3.856) | 0.152*** (3.012) | 0.107*** (3.379) |
| ROA | | 0.141*** (7.161) | 0.115** (2.385) |
| SIZE | | -0.001 (-0.945) | 0.000 (0.152) |
| LEV | | -0.018* (-1.712) | -0.016* (-1.891) |
| MTB | | -0.013*** (-13.327) | -0.013*** (-5.152) |
| AIN | | 0.005** (2.163) | 0.006** (2.680) |
| LIQ | | -0.001 (-0.562) | -0.000 (-0.296) |
| AGE | | 0.010*** (6.273) | 0.009*** (3.967) |
| BODSIZE | | 0.001 (0.871) | 0.000 (0.516) |
| BODIND | | -0.032*** (-2.680) | -0.033* (-1.999) |
| BIG4 | | -0.002 (-0.710) | -0.002 (-0.887) |
| CAP_INTENS | | 0.005*** (7.556) | 0.006*** (4.115) |
| Constant | -0.011*** (-5.290) | 0.013 (0.478) | -0.018 (-0.544) |
| Year Fixed Effect | NO | NO | YES |
| Industry Fixed Effect | NO | NO | YES |
| *Observations* | 1,609 | 1,609 | 1,609 |
| *Adjusted R-squared* | 0.009 | 0.181 | 0.211 |

This table presents the baseline regression results. The dependent variable is earnings management measured by *DACC*, and the main independent variable is environmental innovation (eco-innovation) measured by the environmental innovation score from Refinitiv Eikon. The regression controls for firm-level variables. The robust t-test is reported in parentheses below the coefficients. Standard errors are corrected at the firm level. ***. **, and * denote significance at the 1%, 5%, and 10% levels, respectively.

*4.3. The role of financial constraint as an interaction variable (Hypothesis 2)*

This section examines the moderating effect of financial constraints on the relationship between eco-innovation and earnings management. Following the extant literature, the *WW score* [58] and *sales* [65] proxy financial constraints. Various studies have employed this measure as a proxy for financial constraint [51, 66, 67, 68]. A higher value of the *WW score* reflects more significant financial constraints the firm faces [51]. *The WW score* measure is split into two groups based on the median value. Firms facing high financial constraints receive a value of 1 if their *WW score* exceeds the median value of the *WW score* and vice versa. After the regression estimates for these two groups, as reported in Columns (3) and (4) of Table 6, the findings suggest that the positive relationship between eco-innovation and earnings management is more pronounced for firms facing high financial constraints.

Next, a binary variable is created from the sales value, designating a value of 1 to firms with *sales* below the median, which indicates high financial constraint, and assigning a value of 0 to those reflecting low financial constraint. The findings in Columns (1) and (2) of Table 6 demonstrate a significant correlation between firms with high financial constraints, characterised by *low sales* figures, and their tendency to engage in earnings management while pursuing eco-innovation. Conversely, as evidenced by *high sales*, firms experiencing lower financial constraints are less likely to resort to earnings management practices.

**Table 6.** The moderating role of financial constraint on the relationship between eco-innovation and earnings management.

| *Dependent variable:* | *DACC* | | | |
|---|---|---|---|---|
| Financial Constraint: | *High Sales* | *Low Sales* | *High WW Score* | *Low WW Score* |
|  | (1) | (2) | (3) | (4) |
| EIS | -0.027 (-0.498) | 0.282*** (5.125) | 0.128* (1.842) | 0.095 (1.616) |
| Controls | YES | YES | YES | YES |
| Constant | 0.003 (0.059) | 0.076 (1.534) | 0.013 (0.227) | -0.018 (-0.289) |
| Year Fixed Effect | YES | YES | YES | YES |
| Industry Fixed Effect | YES | YES | YES | YES |
| Observations | 805 | 804 | 805 | 804 |
| Adjusted R-squared | 0.185 | 0.245 | 0.170 | 0.302 |

| *Permutation tests for coef. diff* | p-value < 0.000[1] | p-value = 0.380 |

This table presents the regression results for the moderating effect of financial constraints on the relationship between environmental innovation and earnings management. The dependent variable is earnings management, measured by *DACC*, and the main independent variable is environmental innovation, measured by *EIS*. *Sales* and *WW scores* are the proxies used for financial constraints. The regression controls for firm-level variables. The robust t-test is reported in parentheses next to the coefficients. Standard errors are corrected at the firm level. ***, **, and * denote significance at the 1%, 5%, and 10% levels, respectively. The table presented above demonstrates that organisations experiencing significant financial constraints possess opportunities for earnings management through their environmental innovations, as indicated by the values highlighted in red.

### 4.4. *The role of financial opacity as an interaction variable*

Table 7 reports the regression results, examining the moderating effect of financial opacity on the relationship between Eco-innovation and earnings management. The financial opacity (*FFIN*)[2] variable denotes financial transparency following Dhaliwal et al. [69]. The findings from the analysis suggest that eco-innovative firms with low transparency exhibit strong indications of engaging in earnings management. Conversely, the results of transparent firms with low opacity show no significance, indicating no evidence of earnings management in eco-innovation activities. This finding is consistent with the view that higher transparency in financial information enhances management accountability in financial reporting, thus reducing earnings management practices [70].

**Table 7.** The moderating role of financial opacity on the relationship between eco-innovation and earnings management.

| *Dependent variable* | DACC | |
|---|---|---|
| Financial Opacity: | *High Transparency* | *Low Transparency* |
|  | (1) | (2) |
| EIS | -0.054 (-1.206) | 0.146*** (3.640) |
| *Controls* | YES | YES |
| *Constant* | -0.035 (-0.660) | -0.016 (-0.339) |
| *Year Fixed Effect* | YES | YES |
| *Industry Fixed Effect* | YES | YES |
| *Observations* | 492 | 1,117 |
| *Adjusted R-squared* | 0.212 | 0.222 |

---

[1] When comparing *sales* and *WW scores* as indicators of financial constraints, it is important to note that *sales* have a p-value < 0.000. This suggests that using *sales* as a measure offers a more robust and accurate representation of financial constraints. Therefore, focusing on *sales* would be a more effective approach in this context.

[2] Based on previous research by Bhattacharya et al. [71], firm-level financial transparency is determined by industry and year-adjusted total scaled accruals. The main variable for this model is a scaled accrual, which is an absolute value calculated using the formula ($\Delta CA + \Delta CL + \Delta CASH - \Delta STD + DEP + \Delta TP$) / lag(TA), where $\Delta CA$ represents the change in total current assets, $\Delta CL$ represents the change in total current liabilities, $\Delta CASH$ represents the change in cash, $\Delta STD$ represents the change in the current portion of long-term debt included in total current liabilities - DEP represents depreciation and amortisation expense, $\Delta TP$ represents the change in income taxes payable, lag(TA) represents total assets at the end of the previous year. FFIN takes the value of 1 if a firm has a higher than industry year mean of ACCRUAL and 0 otherwise.

| | |
|---|---|
| *Permutation tests for coef. diff* | p-value = 0.04[3] |

This table presents the regression results for the moderating effect of financial opacity on the relationship between environmental innovation and earnings management. The dependent variable is earnings management, measured by *DACC*, and the main independent variable is environmental innovation, measured by *EIS*. The regression controls for firm-level variables. The robust t-test is reported in parentheses next to the coefficients. Standard errors are corrected at the firm level. ***. **, and * denote significance at the 1%, 5%, and 10% levels, respectively. The table above illustrates that companies with high financial opacity (less transparency) have strong indications of earnings management through their environmental innovation, as indicated by the values highlighted in red.

### 4.5. *The role of ESG performance[4] as an interaction variable*

Table 8 presents the moderating role of *ESG performance* on the relationship between eco-innovation score (*EIS*) and discretionary accruals (*DACC*), our proxy for earnings management. Column (1) reports the estimates for firms classified as firms with high *ESG performance*, while column (2) reflects those with low *ESG performance*.

The findings reveal a pronounced disparity in the correlation between Eco-Innovation Score (*EIS*) and earnings management across ESG categories. Specifically, for firms characterised by low *ESG performance*, the coefficient of *EIS* is positive and statistically significant ($\beta = 0.328$, $t = 4.650$, $p < 0.01$). This suggests that a higher intensity of eco-innovation is associated with elevated levels of discretionary accruals. This result supports the perspective that, without ESG commitments, eco-innovation initiatives may be strategically utilised to obscure financial performance, potentially to obtain external legitimacy or attract sustainability-oriented capital [72, 73].

Conversely, for firms with high *ESG performance*, the coefficient on *EIS* is negative but statistically insignificant ($\beta = -0.029$, $t = -0.590$), indicating no meaningful relationship between eco-innovation intensity and earnings management in this sub-sample. This implies that firms possessing a strong commitment to ESG are less likely to utilise eco-innovation as a means of earnings manipulation, potentially indicating a greater alignment between sustainability initiatives and underlying ethical or governance standards [74].

Importantly, a permutation test assessing the difference in *EIS* coefficients between the two groups yields a highly significant result ($p = 0.000$), reinforcing that ESG performance substantively moderates the *EIS* – earnings management relationship.

**Table 8.** The moderating role of ESG on the relationship between eco-innovation and earnings management.

| *Dependent variable* | DACC | |
|---|---|---|
| ESG Performance: | *High* | *Low* |
| | *ESG Performance* | *ESG Performance* |

---

[3] The permutation test regarding financial opacity indicates that the disparity between companies characterized by high opacity and those with low opacity is statistically significant. This suggests that the influence of financial transparency on these companies is considerable.

[4] We express our sincere gratitude for the invaluable feedback provided by anonymous reviewers, which has surely significantly contributed to the enhancement of the quality of our paper.

|  | (1) | (2) |
|---|---|---|
| *EIS* | -0.029 (-0.590) | 0.328*** (4.650) |
| *Controls* | YES | YES |
| *Constant* | -0.003 (-0.050) | -0.029 (-0.320) |
| *Year Fixed Effect* | YES | YES |
| *Industry Fixed Effect* | YES | YES |
| *Observations* | 805 | 804 |
| *Adjusted R-squared* | 0.243 | 0.228 |
| *Permutation tests for coef. diff* | p-value = 0.000[5] | |

This table presents the regression results for the moderating effect of *ESG Performance* on the relationship between environmental innovation and earnings management. The dependent variable is earnings management, measured by *DACC*, and the main independent variable is environmental innovation, measured by *EIS*. The regression controls for firm-level variables. The robust t-test is reported in parentheses next to the coefficients. Standard errors are corrected at the firm level. ***. **, and * denote significance at the 1%, 5%, and 10% levels, respectively. The table above illustrates that companies with low *ESG performance* have strong indications of earnings management through their environmental innovation, as indicated by the values highlighted in red.

### 4.6. Robustness test
#### 4.6.1. An alternative measure of earnings management

This study employs four alternative earnings management measures to assess the robustness of the main findings. Established in 1991, the Jones model is a widely embraced tool for differentiating accruals and evaluating the discretionary accruals component. This study, therefore, utilises the Jones model [8] to assess the discretionary accruals component, thus, one of the alternative measures of earnings management. Below is the equation to estimate the Jones model of discretionary accruals (*DACC_JONES*):

$$\frac{TA_{it}}{A_{it-1}} = \alpha_1 \left[\frac{1}{A_{it-1}}\right] + \alpha_2 \left[\frac{\Delta REV_{it}}{A_{it-1}}\right] + \alpha_3 \left[\frac{PPE_{it}}{A_{it-1}}\right] + \varepsilon_{it} \quad (5)$$

Where:
$TA_{it}$ = total accrual in year *t* for firm *i*
$\Delta REV_{it}$ = revenues in year *t* less revenue in year *t*-1 for firm *i*
$PPE_{it}$ = gross property, plant, and equipment in year *t* for firm *i*
$A_{it-1}$ = total assets in year *t*–1 for firm *i*
$\varepsilon_{it}$ = error term in year *t* for firm *i*
*i* = 1, . . ., N firm index
*t* = 1, . . ., year index for the years included in the estimation period of firm *i*

Dechow et al. [9] enhanced the Jones [8] model by adjusting the change in revenue with the change in receivables. This modification was based on the premise that manipulating earnings through discretion over revenue recognition for credit sales is more accessible than for cash sales [9]. These adjustments help mitigate the bias towards zero in management's earnings estimates in cases where earnings management has been implemented through revenue management [9]. Below is the equation to estimate the Dechow proxy of earnings management (*DACC_DECHOW*):

$$\frac{TA_{it}}{A_{it-1}} = \alpha_1 \left[\frac{1}{A_{it-1}}\right] + \alpha_2 \left[\frac{\Delta REV_{it} - \Delta REC_{it}}{A_{it-1}}\right] + \alpha_3 \left[\frac{PPE_{it}}{A_{it-1}}\right] + \varepsilon_{it} \qquad (6)$$

---

[5] The permutation test regarding ESG performance indicates that the disparity between companies characterized by high ESG performance and those with low ESG performance is statistically significant. This suggests that the influence of ESG Performance on these companies is considerable.

Where:
*ΔREC$_{it}$*  = receivable in year *t* less receivable in year *t*-1 for firm *i*

The third model, which is widely recognised, is the approach introduced by Kasznik [10]. In this model, Kasznik [10] incorporated operating cash flow due to the negative correlation with total accruals found by Dechow et al. [9]. Below is the equation (*DACC_KASZNIK*):

$$\frac{TA_{it}}{A_{it-1}} = \alpha_1 \left[\frac{1}{A_{it-1}}\right] + \alpha_2 \left[\frac{\Delta REV_{it} - \Delta REC_{it}}{A_{it-1}}\right] + \alpha_3 \left[\frac{PPE_{it}}{A_{it-1}}\right] + \alpha_4 \left[\frac{\Delta CFO_{it}}{A_{it-1}}\right] + \varepsilon_{it} \qquad (7)$$

Where:
*ΔCFO$_{it}$*  = operating cash flow in year *t* less receivable in year *t* - 1 for firm *i*

Lastly, we use the Kothari et al. [11] approach as a proxy for earnings management. This model is distinguished from others because it removes the operating cash flow component in the Kasznik [10] model and integrates company performance, specifically ROA, into the new model. This is because companies identified as exhibiting abnormally high or low levels of earnings management manage more than expected, given their level of performance [11]. Below is the equation (*DACC_KHOTARI*):

$$\frac{TA_{it}}{A_{it-1}} = \alpha_0 + \alpha_1 \left[\frac{1}{A_{it-1}}\right] + \alpha_2 \left[\frac{\Delta REV_{it} - \Delta REC_{it}}{A_{it-1}}\right] + \alpha_3 \left[\frac{PPE_{it}}{A_{it-1}}\right] + \alpha_4 \left[\frac{ROA_{it}}{A_{it-1}}\right] + \varepsilon_{it} \qquad (8)$$

Where:
*ROA$_{it}$* = return on assets in year *t* less receivable in year *t* - 1 for firm *i*

The regression estimate for the four earnings management models as robustness for the relationship between earnings management and eco-innovation is presented in Table 8. A noteworthy observation is the persistently positive and statistically significant correlation of eco-innovation proxies across all models, signifying that an increase in eco-innovation is associated with a greater potential for earnings management, as demonstrated by discretionary accruals in all models.

The coefficients for *DACC_JONES*, *DACC_DECHOW*, *DACC_KASZNIK*, and *DACC_KHOTARI* are 0.098 (p=0.030), 0.098 (p=0.031), 0.081 (p=0.064), and 0.107 (p=0.008), respectively. Notably, the most significant variables that can increase the likelihood of a firm engaging in earnings management are return on assets (*ROA*), the asset income ratio (*AIN*), firm age (*AGE*), and capital intensity (*CAP_INTENS*), as they consistently show positive and significant associations across all models. This indicates that these variables are associated with increased earnings management. Lastly, the consistent significance of *EIS* across all models underscores the strong association between eco-innovation and earnings management.

**Table 8.** Alternative measures of earnings management.

| *Dependent variable* | DACC | | | |
|---|---|---|---|---|
| | *JONES* | *DECHOW* | *KASZNIK* | *KHOTARI* |
| *EIS* | 0.098** (2.584) | 0.098** (2.563) | 0.081* (2.109) | 0.107*** (3.379) |
| *ROA* | 0.398*** (7.547) | 0.403*** (7.780) | 0.462*** (8.925) | 0.115** (2.385) |
| *SIZE* | -0.001 (-0.296) | -0.000 (-0.127) | -0.000 (-0.102) | 0.000 (0.152) |
| *LEV* | 0.005 (0.488) | 0.004 (0.339) | 0.012 (1.213) | -0.016* (-1.891) |
| *MTB* | -0.014*** (-5.330) | -0.014*** (-5.276) | -0.014*** (-5.559) | -0.013*** (-5.152) |
| *AIN* | 0.006** (2.722) | 0.006** (2.704) | 0.006*** (3.932) | 0.006** (2.680) |
| *LIQ* | -0.000 (-0.035) | -0.000 (-0.069) | -0.001 (-0.578) | -0.000 (-0.296) |
| *AGE* | 0.010*** (4.500) | 0.010*** (4.314) | 0.010*** (4.479) | 0.009*** (3.967) |
| *BODSIZE* | 0.001 (0.841) | 0.001 (0.751) | 0.001 (1.233) | 0.000 (0.516) |
| *BODIND* | -0.030 (-1.473) | -0.031 (-1.560) | -0.028 (-1.775) | -0.033* (-1.999) |
| *BIG4* | -0.003 (-0.856) | -0.003 (-0.851) | -0.003 (-0.658) | -0.002 (-0.887) |
| *CAP_INTENS* | 0.006*** (4.401) | 0.006*** (4.131) | 0.006*** (4.425) | 0.006*** (4.115) |
| *Constant* | -0.035 (-0.788) | -0.039 (-0.898) | -0.048 (-1.298) | -0.018 (-0.544) |
| *Year Fixed Effect* | YES | YES | YES | YES |
| *Industry Fixed Effect* | YES | YES | YES | YES |
| *Observations* | 1,609 | 1,609 | 1,609 | 1,609 |
| *Adjusted R-squared* | 0.292 | 0.298 | 0.385 | 0.211 |

Table 8 presents the regression results after employing alternative measures of earnings management. The regression controls for firm-level variables. The robust t-statistics are reported in parentheses below the coefficients. Standard errors are corrected at the firm level. ***, **, and * denote significance at the 1%, 5%, and 10% levels, respectively. Variable definitions are provided in the appendix.

*4.7. Endogeneity test*

Table 9 presents the various checks performed, including entropy balance, propensity score matching, and the Heckman sample selection bias test.

*4.7.1 . Entropy balance*

Entropy balance serves as a valuable method for identifying potential endogeneity test problems. This estimation technique relies on a weighting approach to effectively balance the covariates without requiring adjustments to a propensity model [75]. By utilising entropy balancing, the balance between treatment and control groups can be achieved without excluding any units from either group [76].

Upon reviewing Panel A of Table 9, it becomes apparent that significant differences exist in the means of variables between the treatment and control groups before balancing. This discrepancy has the potential to result in biased treatment effect estimates. However, after applying balancing, the means of variables in the control group closely align with those in the treatment group, indicating successful balancing. After the balancing process, the regression outcomes presented in column (1) of Panel C demonstrate a positive and significant relationship between eco-innovation (*EIS*) and earnings management (*DACC*) after the entropy balance. This finding suggests the potential absence of endogeneity issues based on applying the entropy balancing technique.

*4.7.2. Propensity score matching (PSM)*

Panel B reports the results from Propensity Score Matching (*PSM*) analysis, thus addressing potential endogeneity concerns arising from the difference in the level of eco-innovation among the sampled firms. To address this concern, a logistic regression is first estimated, and the matching techniques are applied, which reduces the sample size. It is observed that despite the reduction in sample size, the matching process allowed for continued significant regression results, especially for eco-innovation and earnings management. All controls and fixed effects were included in the regression. A one-to-one matching without replacement is used with a calliper distance of 0.001. The modest difference between high-intensity and low-intensity groups aligns well with the model's suitability, indicating a promising outcome (see Panel B).

After analysing the results of the *PSM* regression in column (2) of Panel C, it is evident that the coefficient of *EIS* decreased slightly but remains significant even after reducing the number of observations to 706 through matching. *EIS* consistently demonstrates a positive and significant association with the dependent variable *DACC* across various models and adjustments. Moreover, *ROA, AIN, AGE*, and *CAP_INTENS* show a positive correlation with *DACC*, though the strength and significance of this relationship differ. These observations highlight the importance of *EIS* as a robust predictor of the dependent variable *(DACC)*, even after mitigating potential confounding factors through entropy balance and propensity score matching.

### 4.7.3. Heckman Sample Selection Bias Estimation

Panel D reports the two-stage Heckman sample selection bias test [77]. In the first stage, the probit model is applied to identify the factors influencing the likelihood of being in the high eco-innovation *(EIS)* category. Panel D, column 1, the industry average *(IND_AVG)* has a coefficient of 0.316 with a p-value of 0.000, indicating a significant positive effect, suggesting a significant relationship between the *IND_AVG* and *EIS*. The second stage includes the inverse Mills ratio in the primary model. After the estimate, the *EIS* variable shows a coefficient of 0.104 with a p-value of 0.006, demonstrating a significant positive impact on DACC and highlighting its relevance. These findings suggest that the primary results suffer less from endogeneity issues.

**Table 9.** Endogeneity Tests.

**Entropy Balance and Propensity Score Matching**

**Panel A:** Univariate comparison of means between treatment and control groups before and after balancing - Entropy Balance

| | *High Intensity* (*Treated Group*) | *Low Intensity* (*Control Group*) | *High Intensity* (*Treated Group*) | *Low Intensity* (*Control Group*) |
|---|---|---|---|---|
| *Variable* | *Before Balancing* | | *After Balancing* | |
| *ROA* | 0.059 | 0.065 | 0.059 | 0.059 |
| *SIZE* | 21.650 | 20.670 | 21.650 | 21.650 |
| *LEV* | 0.201 | 0.208 | 0.201 | 0.201 |
| *MTB* | 1.642 | 2.007 | 1.642 | 1.642 |
| *AIN* | 0.098 | 0.179 | 0.098 | 0.098 |
| *LIQ* | 1.625 | 1.641 | 1.625 | 1.625 |
| *AGE* | 3.439 | 3.028 | 3.439 | 3.439 |
| *BODSIZE* | 9.186 | 8.423 | 9.186 | 9.186 |
| *BODIND* | 0.643 | 0.588 | 0.643 | 0.643 |
| *BIG4* | 0.630 | 0.654 | 0.630 | 0.630 |
| *CAP_INTENS* | 1.100 | 1.632 | 1.100 | 1.100 |

**Panel B:** Univariate comparison of means between treatment and control groups – PSM

| | *High Intensity* | *Low Intensity* | *Difference of Mean* | |
|---|---|---|---|---|
| *Variable* | (*Treated*) | (*Control*) | *Diff* | *t-value* |
| *ROA* | 0.064 | 0.063 | -0.001 | -0.202 |
| *SIZE* | 21.198 | 21.229 | 0.031 | 0.292 |
| *LEV* | 0.193 | 0.195 | 0.002 | 0.159 |
| *MTB* | 1.864 | 1.752 | -0.112 | -0.865 |
| *AIN* | 0.148 | 0.152 | 0.004 | 0.077 |
| *LIQ* | 1.742 | 1.666 | -0.076 | -0.819 |
| *AGE* | 3.314 | 3.329 | 0.015 | 0.222 |
| *BODSIZE* | 8.722 | 8.832 | 0.110 | 0.736 |
| *BODIND* | 0.617 | 0.609 | -0.008 | -0.837 |
| *BIG4* | 0.677 | 0.649 | -0.028 | -0.795 |
| *CAP_INTENS* | 1.294 | 1.170 | -0.124 | -0.889 |

**Panel C:** Regression Estimate after Entropy Balancing and PSM

| *Dependent variable* | DACC | |
|---|---|---|
| | *Entropy Balancing* | *PSM* |
| | (1) | (2) |
| *EIS* | 0.121** (2.881) | 0.101** (2.374) |
| *ROA* | 0.115** (2.394) | 0.168* (2.228) |
| *SIZE* | -0.000 (-0.142) | 0.003 (1.290) |
| *LEV* | -0.015 (-1.819) | -0.026 (-1.267) |
| *MTB* | -0.013*** (-5.151) | -0.012*** (-3.684) |
| *AIN* | 0.006** (2.708) | 0.010** (2.703) |
| *LIQ* | -0.001 (-0.343) | -0.003 (-0.961) |
| *AGE* | 0.009*** (3.959) | 0.011*** (5.212) |
| *BODSIZE* | 0.000 (0.408) | -0.000 (-0.298) |
| *BODIND* | -0.035** (-2.541) | -0.017 (-0.988) |
| *BIG4* | -0.002 (-0.795) | -0.000 (-0.065) |
| CAP_INTENS | 0.006*** (4.225) | 0.009*** (12.803) |
| Constant | -0.008 (-0.218) | -0.094 (-1.778) |
| Year Fixed Effect | YES | YES |
| Industry Fixed Effect | YES | YES |
| Observations | 1,609 | 706 |
| Adjusted R-squared | 0.211 | 0.220 |

Table 9 Panels A, B, and C present the regression results after propensity score matching analysis (*PSM*) and *entropy balancing*. The dependent variable is earnings management, measured by *DACC*, and the main independent variable is environmental innovation, measured by EIS. The regression controls for firm-level variables. The robust t-statistics are reported in parentheses below the coefficients. Standard errors are corrected at the firm level. ***. **, and * denote significance at the 1%, 5%, and 10% levels, respectively.

**Panel D:** Heckman Test

| | EIS | DACC |
|---|---|---|
| | First Stage | Second Stage |
| VARIABLES | (1) | (2) |
| IND_AVG | 0.316*** (3.390) | |
| EIS | | 0.104*** (3.419) |
| Controls | Yes | Yes |
| Constant | -9.339*** (-4.636) | 0.033 (0.320) |
| Year Fixed Effect | YES | YES |
| Industry Fixed Effect | YES | YES |
| Observations | 1,609 | 1,609 |
| Adjusted R-squared | 0.194 | 0.211 |

Table 9 Panel D presents the regression results after the *Heckman test*. The industry average is used as an instrument in the first stage, with environmental innovation indicators being the dependent variable. In the second stage, the dependent variable is earnings management, measured by *DACC*, and the main independent variable is environmental innovation, measured by *EIS*. The regression controls for firm-level variables. The robust t-statistics are reported in parentheses below the coefficients. Standard errors are corrected at the firm level. ***. **, and * denote significance at the 1%, 5%, and 10% levels, respectively.

## 5. Conclusion

This study explores the intersection of eco-innovation—a critical dimension of environmental sustainability—and earnings management, drawing on a panel of 567 FTSE All-Share firms over the 2014–2022 period. Motivated by growing concerns over corporate transparency and the dual pressures of sustainability and financial performance, this research investigates whether firms engaging in eco-innovation are more prone to earnings management practices. The empirical analysis confirmed a consistent positive relationship between eco-innovation and earnings management, particularly under financial constraints and opacity. These results validate H1–H3 and align with the predictions of legitimacy and agency theories.

The findings reveal a significant positive association between eco-innovation intensity and earnings management, suggesting that firms may strategically employ environmental innovation to meet regulatory and stakeholder expectations and manipulate earnings in pursuit of perceived financial stability (H1). This outcome aligns with prior studies [18], reinforcing that sustainability initiatives coexist with opportunistic financial behaviour. Moreover, the relationship is particularly pronounced in firms experiencing financial constraints (H2) (as measured by low *WW scores* and underwhelming *sales* performance) and those exhibiting financial opacity (H3), proxied by limited financial disclosure (*FFIN*).

To enhance the credibility of the results, the study applies rigorous econometric techniques—entropy balancing, propensity score matching, and the Heckman Test—to address potential endogeneity and selection biases. These approaches strengthen the reliability and generalisability of the core findings.

This research contributes to the broader discourse on how sustainability-linked strategies interact with financial reporting behaviour, offering meaningful implications for regulators, investors, and corporate governance bodies. Specifically, it underscores the importance of critically assessing eco-innovation's dual-use nature as both a signal of environmental responsibility and a potential channel for earnings manipulation. For policymakers, our findings suggest the need for stricter reporting oversight for ESG-active firms to prevent opportunistic behaviour. For investors, the results highlight the importance of monitoring financial transparency alongside ESG disclosures. Corporate boards may also need to align sustainability initiatives with financial reporting integrity frameworks. Theoretically, this study enriches the understanding of how non-financial initiatives like eco-innovation may interplay with opportunistic financial behaviours. Practically, it informs stakeholders about the dual-use nature of ESG activities, where sustainability may co-exist with financial obfuscation.

Despite its contributions, this study is not without limitations. First, the reliance on secondary data limits the ability to capture nuanced managerial motives or firm-level strategic intent behind eco-innovation and earnings management. Second, while our proxies for financial constraints and opacity are consistent with prior literature, they may oversimplify complex realities. Thirdly, while our study uses a model framework to align eco-innovation disclosures and earnings management decisions within the same fiscal period, we acknowledge the potential value of lagged modelling approaches. Due to the temporal structure of ESG data and the theoretical simultaneity of reporting incentives, we did not incorporate lagged predictors in our main model. Furthermore, our concentration on UK firms may restrict the external validity in less-regulated market contexts.

Future research could apply multidimensional indices or incorporate qualitative measures. Furthermore, while methodologically justified, excluding financial and utility sectors may limit the cross-sectoral generalisability. Future research could explore the following areas: Conduct longitudinal firm-level case studies to investigate decision-making dynamics; Extend the analysis to encompass other national contexts to evaluate regulatory influences; Examine how various components of ESG -such as social and governance- interact with earnings management; and employ qualitative interviews with executives to uncover managerial intentions behind earnings management within ESG contexts. Future research may consider applying dynamic panel data techniques, such as System GMM, or testing multi-period lags where data availability permits to explore causality in this context further.

Author Contributions
Conceptualisation, P.E.S., M.A. and P.K.; methodology, P.E.S., M.A. and P.K.; software, P.E.S.; validation, M.A. and P.K.; formal analysis, P.E.S.; investigation, P.E.S.; data curation, P.E.S.; writing—original draft preparation, P.E.S.; writing—review and editing, P.E.S., M.A. and P.K.; supervision, M.A. and P.K.; project administration, M.A. and P.K.; funding acquisition, P.E.S. All authors have read and agreed to the published version of the manuscript.

Data availability
The raw data supporting the conclusions of this article will be made available by the authors on request.

Conflict of Interest
The authors declare no conflict of interest.

Acknowledgements: This research was funded by the Centre for Higher Education Funding and Assessment (PPAPT), the Ministry of Education, Science, and Technology of the Republic of Indonesia (BPI), Ref. Number 2944/BPPT/BPI.LG/IV/2024, and the Indonesian Endowment Fund for Education (LPDP) in the form of a full scholarship for a PhD student of P.E.S.